\documentclass[%
reprint,
superscriptaddress,
showpacs,
 amsmath,amssymb,
 aps,
 prl,
]{revtex4-1}

\usepackage{graphicx}
\usepackage{dcolumn}
\usepackage{bm}
\usepackage{hyperref}
\usepackage{natbib}
\usepackage{multirow}
\usepackage{color}

\begin{document}

\preprint{APS/123-QED}

\title{Search for solar axions by Primakoff effect with the full dataset of the CDEX-1B Experiment}

\author{L.~T.~Yang}
\affiliation{Key Laboratory of Particle and Radiation Imaging (Ministry of Education) and Department of Engineering Physics, Tsinghua University, Beijing 100084}
\author{S.~K.~Liu}
\email{Corresponding author: liusk@scu.edu.cn}
\affiliation{College of Physics, Sichuan University, Chengdu 610065}
\author{Q.~Yue}
\email{Corresponding author: yueq@mail.tsinghua.edu.cn}
\affiliation{Key Laboratory of Particle and Radiation Imaging (Ministry of Education) and Department of Engineering Physics, Tsinghua University, Beijing 100084}

\author{K.~J.~Kang}
\affiliation{Key Laboratory of Particle and Radiation Imaging (Ministry of Education) and Department of Engineering Physics, Tsinghua University, Beijing 100084}
\author{Y.~J.~Li}
\affiliation{Key Laboratory of Particle and Radiation Imaging (Ministry of Education) and Department of Engineering Physics, Tsinghua University, Beijing 100084}

\author{H.~P.~An}
\affiliation{Key Laboratory of Particle and Radiation Imaging (Ministry of Education) and Department of Engineering Physics, Tsinghua University, Beijing 100084}
\affiliation{Department of Physics, Tsinghua University, Beijing 100084}

\author{Greeshma~C.}
\altaffiliation{Participating as a member of TEXONO Collaboration}
\affiliation{Institute of Physics, Academia Sinica, Taipei 11529}

\author{J.~P.~Chang}
\affiliation{NUCTECH Company, Beijing 100084}

\author{Y.~H.~Chen}
\affiliation{YaLong River Hydropower Development Company, Chengdu 610051}
\author{J.~P.~Cheng}
\affiliation{Key Laboratory of Particle and Radiation Imaging (Ministry of Education) and Department of Engineering Physics, Tsinghua University, Beijing 100084}
\affiliation{College of Nuclear Science and Technology, Beijing Normal University, Beijing 100875}
\author{W.~H.~Dai}
\affiliation{Key Laboratory of Particle and Radiation Imaging (Ministry of Education) and Department of Engineering Physics, Tsinghua University, Beijing 100084}
\author{Z.~Deng}
\affiliation{Key Laboratory of Particle and Radiation Imaging (Ministry of Education) and Department of Engineering Physics, Tsinghua University, Beijing 100084}
\author{C.~H.~Fang}
\affiliation{College of Physics, Sichuan University, Chengdu 610065}
\author{X.~P.~Geng}
\affiliation{Key Laboratory of Particle and Radiation Imaging (Ministry of Education) and Department of Engineering Physics, Tsinghua University, Beijing 100084}
\author{H.~Gong}
\affiliation{Key Laboratory of Particle and Radiation Imaging (Ministry of Education) and Department of Engineering Physics, Tsinghua University, Beijing 100084}
\author{Q.~J.~Guo}
\affiliation{School of Physics, Peking University, Beijing 100871}
\author{T.~Guo}
\affiliation{Key Laboratory of Particle and Radiation Imaging (Ministry of Education) and Department of Engineering Physics, Tsinghua University, Beijing 100084}
\author{X.~Y.~Guo}
\affiliation{YaLong River Hydropower Development Company, Chengdu 610051}
\author{L.~He}
\affiliation{NUCTECH Company, Beijing 100084}
\author{J.~R.~He}
\affiliation{YaLong River Hydropower Development Company, Chengdu 610051}
\author{J.~W.~Hu}
\affiliation{Key Laboratory of Particle and Radiation Imaging (Ministry of Education) and Department of Engineering Physics, Tsinghua University, Beijing 100084}
\author{H.~X.~Huang}
\affiliation{Department of Nuclear Physics, China Institute of Atomic Energy, Beijing 102413}
\author{T.~C.~Huang}
\affiliation{Sino-French Institute of Nuclear and Technology, Sun Yat-sen University, Zhuhai 519082}
\author{L.~Jiang}
\affiliation{Key Laboratory of Particle and Radiation Imaging (Ministry of Education) and Department of Engineering Physics, Tsinghua University, Beijing 100084}
\author{S.~Karmakar}
\altaffiliation{Participating as a member of TEXONO Collaboration}
\affiliation{Institute of Physics, Academia Sinica, Taipei 11529}

\author{H.~B.~Li}
\altaffiliation{Participating as a member of TEXONO Collaboration}
\affiliation{Institute of Physics, Academia Sinica, Taipei 11529}
\author{H.~Y.~Li}
\affiliation{College of Physics, Sichuan University, Chengdu 610065}
\author{J.~M.~Li}
\affiliation{Key Laboratory of Particle and Radiation Imaging (Ministry of Education) and Department of Engineering Physics, Tsinghua University, Beijing 100084}
\author{J.~Li}
\affiliation{Key Laboratory of Particle and Radiation Imaging (Ministry of Education) and Department of Engineering Physics, Tsinghua University, Beijing 100084}
\author{M.~C.~Li}
\affiliation{YaLong River Hydropower Development Company, Chengdu 610051}
\author{Q.~Y.~Li}
\affiliation{College of Physics, Sichuan University, Chengdu 610065}
\author{R.~M.~J.~Li}
\affiliation{College of Physics, Sichuan University, Chengdu 610065}
\author{X.~Q.~Li}
\affiliation{School of Physics, Nankai University, Tianjin 300071}
\author{Y.~L.~Li}
\affiliation{Key Laboratory of Particle and Radiation Imaging (Ministry of Education) and Department of Engineering Physics, Tsinghua University, Beijing 100084}
\author{Y.~F.~Liang}
\affiliation{Key Laboratory of Particle and Radiation Imaging (Ministry of Education) and Department of Engineering Physics, Tsinghua University, Beijing 100084}
\author{B.~Liao}
\affiliation{College of Nuclear Science and Technology, Beijing Normal University, Beijing 100875}
\author{F.~K.~Lin}
\altaffiliation{Participating as a member of TEXONO Collaboration}
\affiliation{Institute of Physics, Academia Sinica, Taipei 11529}
\author{S.~T.~Lin}
\affiliation{College of Physics, Sichuan University, Chengdu 610065}
\author{J.~X.~Liu}
\affiliation{Key Laboratory of Particle and Radiation Imaging (Ministry of Education) and Department of Engineering Physics, Tsinghua University, Beijing 100084}
\author{Y.~D.~Liu}
\affiliation{College of Nuclear Science and Technology, Beijing Normal University, Beijing 100875}
\author{Y.~Liu}
\affiliation{College of Physics, Sichuan University, Chengdu 610065}
\author{Y.~Y.~Liu}
\affiliation{College of Nuclear Science and Technology, Beijing Normal University, Beijing 100875}
\author{H.~Ma}
\affiliation{Key Laboratory of Particle and Radiation Imaging (Ministry of Education) and Department of Engineering Physics, Tsinghua University, Beijing 100084}
\author{Y.~C.~Mao}
\affiliation{School of Physics, Peking University, Beijing 100871}
\author{Q.~Y.~Nie}
\affiliation{Key Laboratory of Particle and Radiation Imaging (Ministry of Education) and Department of Engineering Physics, Tsinghua University, Beijing 100084}
\author{H.~Pan}
\affiliation{NUCTECH Company, Beijing 100084}
\author{N.~C.~Qi}
\affiliation{YaLong River Hydropower Development Company, Chengdu 610051}
\author{J.~Ren}
\affiliation{Department of Nuclear Physics, China Institute of Atomic Energy, Beijing 102413}
\author{X.~C.~Ruan}
\affiliation{Department of Nuclear Physics, China Institute of Atomic Energy, Beijing 102413}
\author{M.~B.~Shen}
\affiliation{YaLong River Hydropower Development Company, Chengdu 610051}
\author{M.~K.~Singh}
\altaffiliation{Participating as a member of TEXONO Collaboration}
\affiliation{Institute of Physics, Academia Sinica, Taipei 11529}
\affiliation{Department of Physics, Banaras Hindu University, Varanasi 221005}
\author{T.~X.~Sun}
\affiliation{College of Nuclear Science and Technology, Beijing Normal University, Beijing 100875}
\author{W.~L.~Sun}
\affiliation{YaLong River Hydropower Development Company, Chengdu 610051}
\author{C.~J.~Tang}
\affiliation{College of Physics, Sichuan University, Chengdu 610065}
\author{Y.~Tian}
\affiliation{Key Laboratory of Particle and Radiation Imaging (Ministry of Education) and Department of Engineering Physics, Tsinghua University, Beijing 100084}
\author{G.~F.~Wang}
\affiliation{College of Nuclear Science and Technology, Beijing Normal University, Beijing 100875}
\author{J.~Z.~Wang}
\affiliation{Key Laboratory of Particle and Radiation Imaging (Ministry of Education) and Department of Engineering Physics, Tsinghua University, Beijing 100084}
\author{L.~Wang}
\affiliation{Department of  Physics, Beijing Normal University, Beijing 100875}
\author{Q.~Wang}
\affiliation{Key Laboratory of Particle and Radiation Imaging (Ministry of Education) and Department of Engineering Physics, Tsinghua University, Beijing 100084}
\affiliation{Department of Physics, Tsinghua University, Beijing 100084}
\author{Y.~F.~Wang}
\affiliation{Key Laboratory of Particle and Radiation Imaging (Ministry of Education) and Department of Engineering Physics, Tsinghua University, Beijing 100084}
\author{Y.~X.~Wang}
\affiliation{School of Physics, Peking University, Beijing 100871}
\author{H.~T.~Wong}
\altaffiliation{Participating as a member of TEXONO Collaboration}
\affiliation{Institute of Physics, Academia Sinica, Taipei 11529}

\author{Y.~C.~Wu}
\affiliation{Key Laboratory of Particle and Radiation Imaging (Ministry of Education) and Department of Engineering Physics, Tsinghua University, Beijing 100084}
\author{H.~Y.~Xing}
\affiliation{College of Physics, Sichuan University, Chengdu 610065}
\author{K.~Z.~Xiong}
\affiliation{YaLong River Hydropower Development Company, Chengdu 610051}
\author{R. Xu}
\affiliation{Key Laboratory of Particle and Radiation Imaging (Ministry of Education) and Department of Engineering Physics, Tsinghua University, Beijing 100084}
\author{Y.~Xu}
\affiliation{School of Physics, Nankai University, Tianjin 300071}
\author{T.~Xue}
\affiliation{Key Laboratory of Particle and Radiation Imaging (Ministry of Education) and Department of Engineering Physics, Tsinghua University, Beijing 100084}
\author{Y.~L.~Yan}
\affiliation{College of Physics, Sichuan University, Chengdu 610065}
\author{N.~Yi}
\affiliation{Key Laboratory of Particle and Radiation Imaging (Ministry of Education) and Department of Engineering Physics, Tsinghua University, Beijing 100084}
\author{C.~X.~Yu}
\affiliation{School of Physics, Nankai University, Tianjin 300071}
\author{H.~J.~Yu}
\affiliation{NUCTECH Company, Beijing 100084}
\author{M.~Zeng}
\affiliation{Key Laboratory of Particle and Radiation Imaging (Ministry of Education) and Department of Engineering Physics, Tsinghua University, Beijing 100084}
\author{Z.~Zeng}
\affiliation{Key Laboratory of Particle and Radiation Imaging (Ministry of Education) and Department of Engineering Physics, Tsinghua University, Beijing 100084}
\author{B.~T.~Zhang}
\affiliation{Key Laboratory of Particle and Radiation Imaging (Ministry of Education) and Department of Engineering Physics, Tsinghua University, Beijing 100084}
\author{F.~S.~Zhang}
\affiliation{College of Nuclear Science and Technology, Beijing Normal University, Beijing 100875}
\author{L.~Zhang}
\affiliation{College of Physics, Sichuan University, Chengdu 610065}
\author{P.~Zhang}
\affiliation{YaLong River Hydropower Development Company, Chengdu 610051}
\author{Z.~H.~Zhang}
\affiliation{Key Laboratory of Particle and Radiation Imaging (Ministry of Education) and Department of Engineering Physics, Tsinghua University, Beijing 100084}
\author{Z.~Y.~Zhang}
\affiliation{Key Laboratory of Particle and Radiation Imaging (Ministry of Education) and Department of Engineering Physics, Tsinghua University, Beijing 100084}
\author{J.~Z.~Zhao}
\affiliation{Key Laboratory of Particle and Radiation Imaging (Ministry of Education) and Department of Engineering Physics, Tsinghua University, Beijing 100084}
\author{K.~K.~Zhao}
\affiliation{College of Physics, Sichuan University, Chengdu 610065}
\author{M.~G.~Zhao}
\affiliation{School of Physics, Nankai University, Tianjin 300071}

\author{J.~F.~Zhou}
\affiliation{YaLong River Hydropower Development Company, Chengdu 610051}
\author{Z.~Y.~Zhou}
\affiliation{Department of Nuclear Physics, China Institute of Atomic Energy, Beijing 102413}
\author{J.~J.~Zhu}
\affiliation{College of Physics, Sichuan University, Chengdu 610065}

\collaboration{CDEX Collaboration}
\noaffiliation
\date{\today}

\begin{abstract}
We present the first limit on $g_{A\gamma}$ coupling constant using the Bragg-Primakoff conversion based on an exposure of 1107.5 kg days of data from the CDEX-1B experiment at the China Jinping Underground Laboratory. The data are consistent with the null signal hypothesis, and no excess signals are observed. Limits of the coupling $g_{A\gamma}<2.08\times10^{-9}$ GeV$^{-1}$ (95\% C.L.) are derived for axions with mass up to 100 eV/$c^2$. Within the hadronic model of KSVZ, our results exclude axion mass $>5.3~\rm{eV}/c^2$ at 95\% C.L.

\end{abstract}

\maketitle

\section{I. Introduction}
Quantum chromodynamics (QCD) is a charge-parity (CP) violating theory containing the $\Theta$, which could cause measurable CP-violating effects, such as a high neutron electric dipole moment. However, the experimental upper bound is about 10$^{10}$ times larger, resulting in an extremely unnatural $\Theta$. To solve this ``strong CP problem", Peccei and Quinn introduced a new spontaneously broken symmetry that causes a strong CP violation to vanish dynamically~\cite{Axion_PRL_1977,Axion_PRD_1977}. Subsequently, it was demonstrated that the Peccei-Quinn mechanism could generate a new Nambu-Goldston boson called axion~\cite{Axion_PRL_1978,Axion_PRL_1978_Wilczek}. Subsequent investigations swiftly discredited the original axion associated with the electroweak scale. However, ``invisible" axion models such as the nonhadronic DFSZ model~\cite{DINE1981199,1980_Zhitniskiy} and the hadronic KSVZ model~\cite{Kim_PRL1979,SHIFMAN1980493} emerging from a higher symmetry-breaking energy scale are still allowed.
Axions or axion-like particles, the light pseudoscalar bosons, are also one of the leading candidates for dark matter~\cite{JPreskill_1983,LFAbbott_1983,MDine_1983}, which may have model-dependent couplings to photons ($g_{A\gamma}$), electron ($g_{Ae}$), and nucleons ($g_{AN}$). In the hadronic model, axions can be coupled to the new, heavy quarks and do not interact with ordinary quarks and leptons at the tree level, leading to a significant suppression of $g_{Ae}$.

Based on the assumption of the axion hadronic model, we focus on the searches for solar axion originating from the Primakoff production, $\gamma+Q \to Q+A$ ($Q$ stands for charged particles), which is one of the primary mechanisms leading to axion production in the Sun. In contrast, some experiments utilize the Primakoff effect $A+Q \to Q+\gamma$ to search for solar axions. In addition, these crystal experiments can detect solar axions through the Primakoff and Bragg diffraction effects. Similar to the Bragg diffraction of X-rays, the Bragg diffraction effect of axions will significantly enhance detection.
The axion-photon coupling $g_{A\gamma}$ from this process is independent of the axion mass. These constraints on $g_{A\gamma}$ have the most stringent laboratory bounds above 1.17 eV/$c^2$~\cite{IRASTORZA2018} compared to the helioscope~\cite{Arik-CAST:PRL2014,CAST:Nature2017} and microwave cavity experiments~\cite{ADMX:PRL2018}.

The SOLAX experiment was a pioneer in searching for solar axions using the Bragg-Primakoff conversion. The SOLAX team developed the first detector-signature phenomenology upon which the Bragg scattering analysis is based~\cite{Paschos:PLB1994,CRESWICK:PLB1998}. Similar to SOLAX~\cite{SOLAX:PRL1998}, the COSME experiment used a Ge detector for their searches~\cite{MORALES:AstroPhys2002}. Both the CDMS~\cite{CDMS_PRL2009} and EDELWEISS~\cite{Armengaud:JCAP2013} experiments used Ge detectors configured as bolometers. The CDMS result is noteworthy because of the well-known orientation of the crystal axis with respect to the Sun. The Majorana Demonstrator experiment has searched for solar axions with a set of $^{76}$Ge-enriched high purity germanium detectors using a 33 kg-yr exposure and gave the best laboratory-based limits on the axion-photon coupling between 1~eV/c$^2$ and 100~eV/c$^2$ to date as $g_{A\gamma}<1.45\times10^{-9}$ GeV$^{-1}$~\cite{Majorana_PRL2022}. The DAMA experiment~\cite{DAMA:PLB2001} used NaI crystals. Although the DAMA measurement has far more extensive exposure than Ge experiments, its energy resolution is low. The TEXONO experiment sought coherently interacting axions coming from a reactor~\cite{TEXONO_PRD_2007}.

The China Dark Matter Experiment (CDEX)~\cite{CDEX_introduction,CDEX_1kg_2014,CDEX_1kg_2016,Yang:2018b,CDEX-AM:PRL2019,Jiang:2018,cdex10_tech,cdexmigdal,cdexdarkphoton,CDEX_axion_2017,CDEX-1B_axion_2019,cdex10_eft} pursues direct searches of light dark matter and studies of neutrinoless double-beta decay of $^{76}$Ge~\cite{cdex0vbb2017} toward the goal of a ton-scale germanium detector array at the China Jinping Underground Laboratory (CJPL)~\cite{CJPL_annurev2017}. CJPL is located in the Jinping traffic tunnel in the Sichuan province of China, with a vertical rock overburden of more than 2400 m, giving rise to  a measured muon flux of 61.7 y$^{-1}$ m$^{-2}$~\cite{WUYC_CPC_2013}. The solar axions and ALPs searches of coupling constants of axion-electrons and axion-nucleons have been reported by the CDEX-1A~\cite{CDEX_axion_2017} and updated CDEX-1B~\cite{CDEX-1B_axion_2019} experiments in stage CDEX-1, both with one crystal of the p-type point-contact germanium ($p$PCGe) of $\mathcal{O}$(1 kg) mass.
In this paper, we present the first results of the $g_{A\gamma}$ coupling constant through the Bragg-Primakoff conversion from the CDEX-1B experiment, using an exposure of 1107.5 kg days of data ~\cite{CDEX-AM:PRL2019}.

\section{II. Axion searches with CDEX-1B}

\subsection{A. CDEX-1B setup and overview}
The CDEX-1B experiment~\cite{Yang:2018b,CDEX-AM:PRL2019,cdexmigdal,CDEX-1B_axion_2019},the second phase of the CDEX-1 experiment, was conducted at the CJPL. Its passive shielding system includes, from outside to inside, 1m of polyethylene, 20 cm of lead, 20 cm of borated polyethylene, and 20 cm of oxygen-free high conductivity (OFHC) copper. The anti-Compton detector uses a well-shaped cylindrical NaI(Tl) crystal surrounding the PPCGe detector. To establish a positive pressure and further prevent radon intrusion, nitrogen gas evaporated from liquid nitrogen is forced into an acrylic box, which encloses the OFHC copper shield.

CDEX-1B germanium detector has a target mass of 939 g and a dead layer of 0.88 $\pm$ 0.12 mm~\cite{C1B_deadlayer_2017}. The vertical axis of the cryostat and OFHC copper end cap is aligned with the [001] axis of the germanium crystal, with a precision within 1 degree. However, the [010] and [001] axes relative to the cryostat were not measured.

The DAQ system received signals from the $p^{+}$ point contact electrode of CDEX-1B, which were fed into a pulsed reset preamplifier. Five identical output signals of the preamplifier were further processed and digitized. Two were distributed into 6 $\mu$s and 12 $\mu$s shaping amplifiers for the 0$-$12 keV energy range. These two channels were used for energy calibration and signal and noise discrimination. The third channel was loaded with a timing amplifier to measure the rise time of signals within an energy range of 0$-$12 keV, which can be used for bulk or surface events discrimination. The remaining two were loaded with low-gain shaping and timing amplifiers, aiming for a high energy range for background understanding. To estimate the dead time of the DAQ system and selection efficiencies uncorrelated with energy, random trigger events were recorded every 20 seconds. The output signals of the above amplifiers were digitized by the 14-bit, 100-MHz flash analog-to-digital converters (FADC). The time tag register provides the event time information with a time resolution of 20 ns.

\subsection{B. Solar axion sources}

The Sun is a potential source of axions~\cite{GGRaffelt_1986,MPospelov_2008,PGondolo_2009,AVDerbin_2011,JRedondo_2013} because of the high abundance of photons and strong electromagnetic fields in the Sun. Axion can be efficiently produced by the inverse Primakoff conversion in the fluctuating electric field of the plasma, which allows for the conversion of the photon into an axion: $\gamma \to A$~\cite{DADicus_1978,GGRaffelt_1986,GGRaffelt_1988}. 

\begin{figure}[!htbp]
\includegraphics[width=\linewidth]{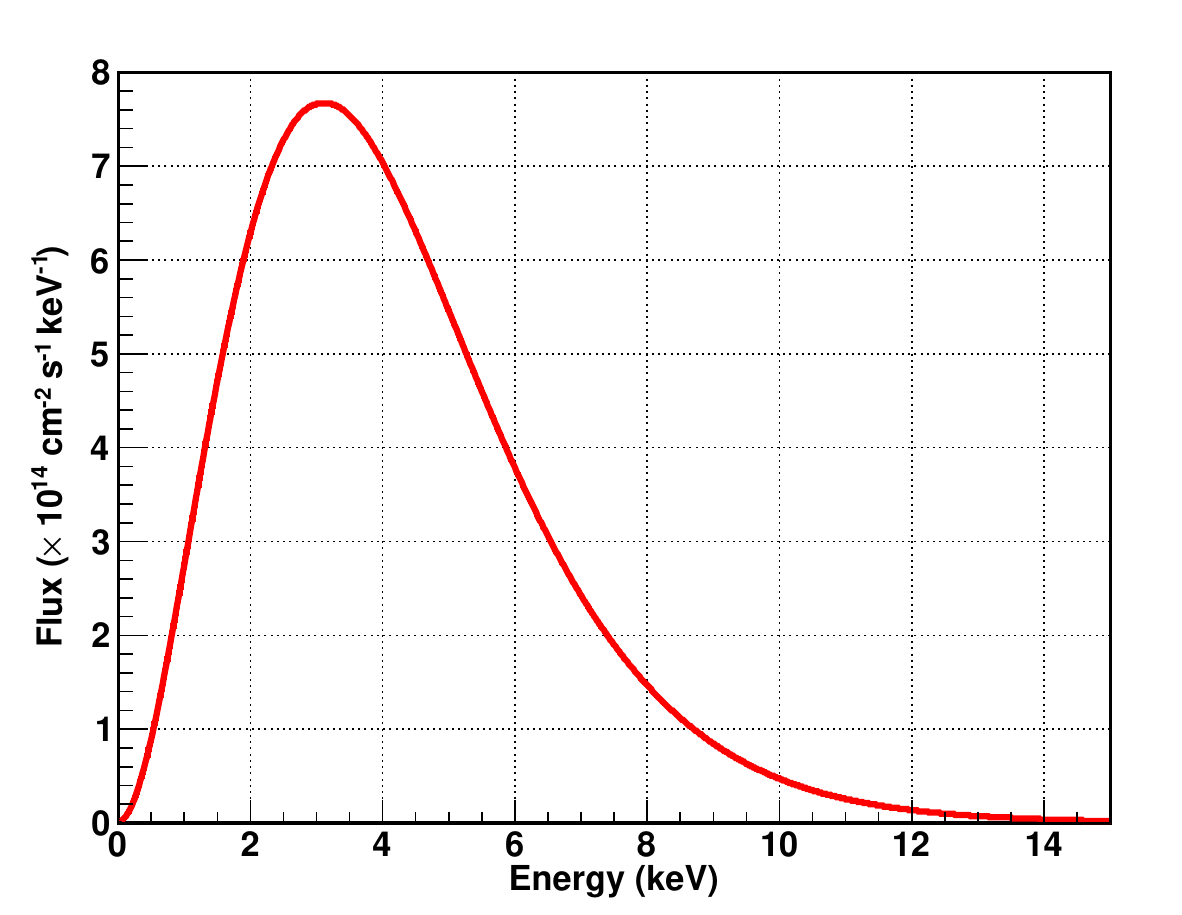}
\caption{
Axion flux spectrum escaping from the Sun for a coupling constant of $g_{A\gamma}=10^{-8} ~\text{GeV}^{-1}$.
}
\label{fig:axion_flux}
\end{figure}

The expected flux of solar axions on Earth can be calculated assuming the standard solar model ~\cite{CDMS_PRL2009,Bahcall_PRL2004} and coupling to the keV-scale blackbody photons in the core region of the Sun, which create a flux of $\mathcal{O}$(keV) axions on Earth. The expected solar Primakoff axion flux is well approximated by~\cite{Andriamonje_2007,Arik_2009}
\begin{equation}
\begin{aligned}
\frac{d\Phi}{dE} = \frac{6.02\times 10^{14}}{\text{cm}^2 \text{ keV s}} (\frac{g_{A\gamma}\times 10^8}{\text{GeV}^{-1}})^2 E^{2.481} e^{-E/1.205},
\end{aligned}
\label{eq:axion_flux}
\end{equation}
where $E$ is the energy of the axion in keV and $g_{A\gamma}$ is the axion-photon coupling constant. The axion flux shown in Fig.~\ref{fig:axion_flux}, whose intensity varies as $g_{A\gamma}^2$, has a continuous spectrum with a peak near 4 keV and falls off exponentially beyond about 10 keV. 

\subsection{C. Experimental Signatures}

The solar axion produced by the inverse Primakoff conversion can be detected again by the Primakoff effect, which shows that axions can pass in the proximity of the atomic nuclei of the crystal, where the intense electric field can trigger their conversion into photons. In this process, the energy of the outgoing photon is equal to that of the incoming axion. The axion-photon effective Lagrangian is 
\begin{eqnarray}
\label{eq_axion_photon_L}
\mathcal{L} = - \frac{1}{4}g_{A\gamma}F^{\mu \nu}\widetilde{F}_{\mu \nu}\phi_{A}=g_{A\gamma}\emph{\textbf{E}}\cdot \emph{\textbf{B}} \phi_A,
\end{eqnarray}
where $F^{\mu \nu}$ is the electromagnetic field tensor, $\widetilde{F}_{\mu \nu}$ its dual, and $\emph{\textbf{E}}$ and $\emph{\textbf{B}}$ are the electric and magnetic fields, respectively~\cite{Raffelt_2007}. $\phi_{A}$ the pseudoscalar axion field and $g_{A\gamma}$ the axion-photon coupling constant. 
Within standard axion models, this coupling can be written as~\cite{DAMA:PLB2001}:
\begin{eqnarray}
\label{eq_g_a_gamma}
g_{A\gamma} \simeq 0.19 \frac{m_A}{\text{eV}}\left| \frac{E}{N}-\frac{2(4+z)}{3(1+z)}\right| 10^{-9} \text{ GeV}^{-1},
\end{eqnarray}
where $m_A$ is the axion mass and $E/N$ is the Peccei-Quinn symmetry anomaly, in particular, $E/N=8/3$ or $E/N = 0$ for the DFSZ~\cite{DINE1981199,1980_Zhitniskiy} and KSVZ~\cite{Kim_PRL1979,SHIFMAN1980493} models, respectively. $z\equiv m_u/m_d \simeq 0.56$ is the mass ratio of the lightest up and down quarks, noting that $z$ still suffers significant uncertainties~\cite{MOROI199869,Workman_2022ynf}.

In addition, a coherent effect can be produced when the Bragg condition is fulfilled ($2d \sin\theta = n\lambda$), similar to the Bragg reflection of X-rays, which leads to a strong enhancement of the signal. Because this enhancement depends on the direction of the incoming axion with respect to the planes of the crystal lattice, i.e., the detection rates in a certain energy window vary with the relative orientations of the crystal and the Sun, the Bragg diffraction condition creates a unique ``fingerprint" according to the location, orientation, time, and energy range~\cite{WBuchmuller_1990,Paschos:PLB1994,CRESWICK:PLB1998}.

Following the derivation in Refs~\cite{CRESWICK:PLB1998,SOLAX:PRL1998,XuWQ_2017}, the total rate of axion-converted photons with incoming axion energy $E$ in a Ge crystal of volume $V$ can be expressed,
\begin{align}
\frac{dR}{dE}(\hat{k},E) = 2\hbar c \frac{V}{v_c^2}\sum\limits_{\boldsymbol{G}}{\frac{d\Phi}{dE} \frac{\vert S(\boldsymbol{G}) \vert^2}{\vert \boldsymbol{G} \vert^2}} \frac{d\sigma}{d\Omega} \times \delta(E-\frac{\hbar c \vert \boldsymbol{G} \vert^2}{2\hat{k}\cdot\boldsymbol{G}}),
\label{eq:dR_dE}
\end{align}
where $v_c$ is the volume of a unit cell; $\hat{k}\equiv \boldsymbol{k}/\vert \boldsymbol{k} \vert$ represents the direction of the solar axion flux; $\boldsymbol{G}=2\pi(h,k,l)/a_0$ is a reciprocal lattice vector, $h,k,l$ are integers and each $(h,k,l)$ represents a family of planes of the crystal. $d\Phi/dE$ is defined in Eq.~\ref{eq:axion_flux} evaluated at the axion energy of $\hbar c \vert \boldsymbol{G} \vert^2 / (2\widehat{k}\cdot\boldsymbol{G})$. $S(\boldsymbol{G})$ is the crystal structure function, which for germanium is defined by~\cite{CRESWICK:PLB1998}:
\begin{eqnarray}
\begin{aligned} 
\label{eq_g_S_G}
S(\boldsymbol(G))= & [1+e^{i\pi(h+k+l)/2}] \times \\
& [1+e^{i\pi(h+k)}+e^{i\pi(h+l)}+e^{i\pi(k+l)}].
\end{aligned}
\end{eqnarray}

The differential cross section for Primakoff conversion on an atom is given~\cite{CRESWICK:PLB1998}:
\begin{equation} 
\frac{d\sigma}{d\Omega} = \frac{g_{A\gamma}^2}{16\pi^2}F_A^2(2\theta)\sin^2(2\theta),
\end{equation}
where $2\theta$ is the scattering angle. $F_A(2\theta)$ is the form factor of the screened Coulomb field of the nucleus.
The remaining delta function results from the Bragg condition expressed in energy. This rate depends on the fourth power of $g_{A\gamma}$.

The expected axion event rate at measurable energy $E_{ee}$ (``ee" represents electron equivalent energy) is obtained by the convolution of the axion-converted photon rate (defined in Eq.~\ref{eq:dR_dE}) and the energy resolution of the detector:
\begin{eqnarray}
\label{eq1}
R(\hat{k}, E_{ee})=\int{dE \frac{dR}{dE}(\hat{k},E) \times\frac{1}{\sqrt{2\pi}\sigma}e^{-(E_{ee}-E)^{2}/2\sigma^{2}}}.
\end{eqnarray}

The $\hat{k}$ varies with the instantaneous time $t$, i.e., $R(\hat{k}(t),E_{ee})\equiv R(t,E_{ee})$. 
In particular, the (001) axis of our crystal is known, aligned with the vertical axis of the cryostat. A spherical coordinate system can be established for the laboratory, with the crystal (001) axis as the $z$ direction and the unknown absolute azimuthal angle $\phi$ of horizontal crystal axes. Therefore, $\hat k$ is also relative to $\phi$. The standard deviation of detector energy resolution, $\sigma$, is 83 eV at 10.37 keV~\cite{Yang:2018b,CDEX-1B_axion_2019}.

\begin{figure}[!htbp]
\includegraphics[width=\linewidth]{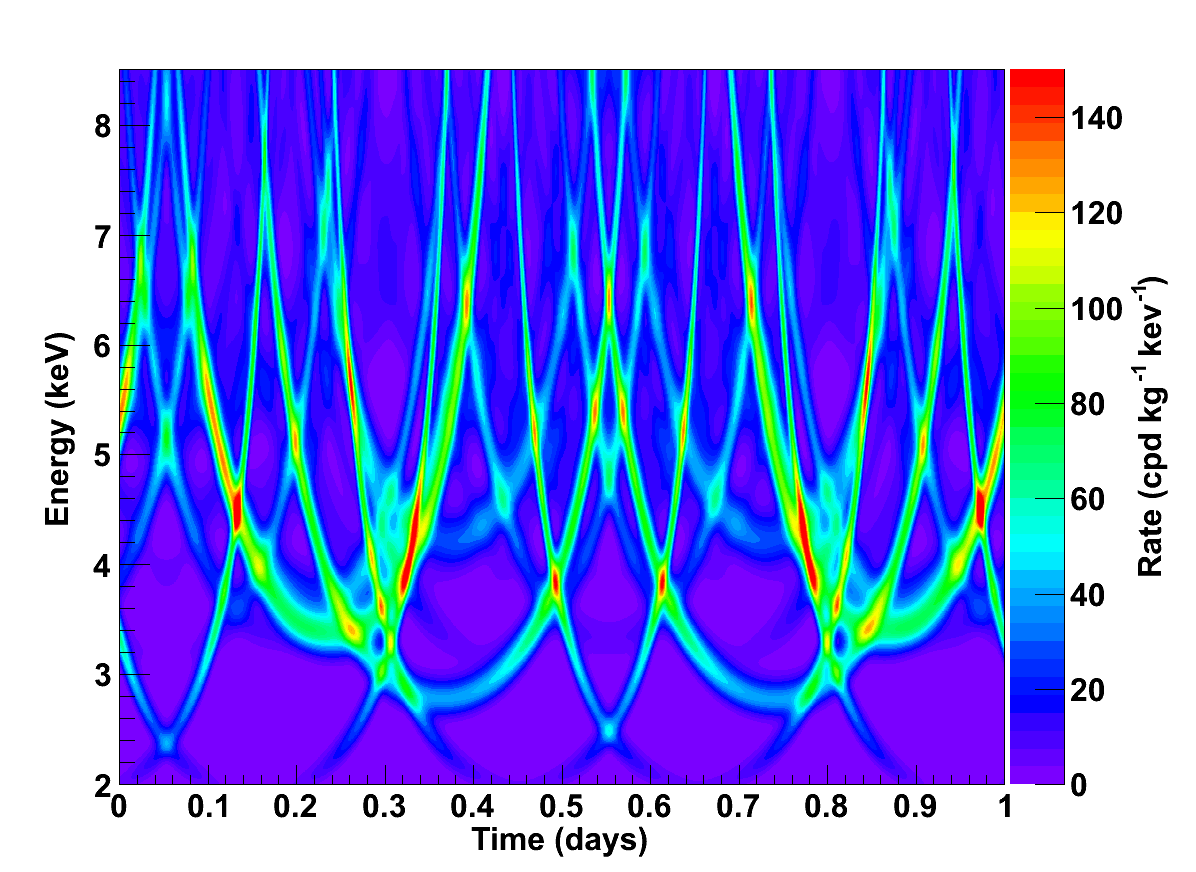}
\includegraphics[width=\linewidth]{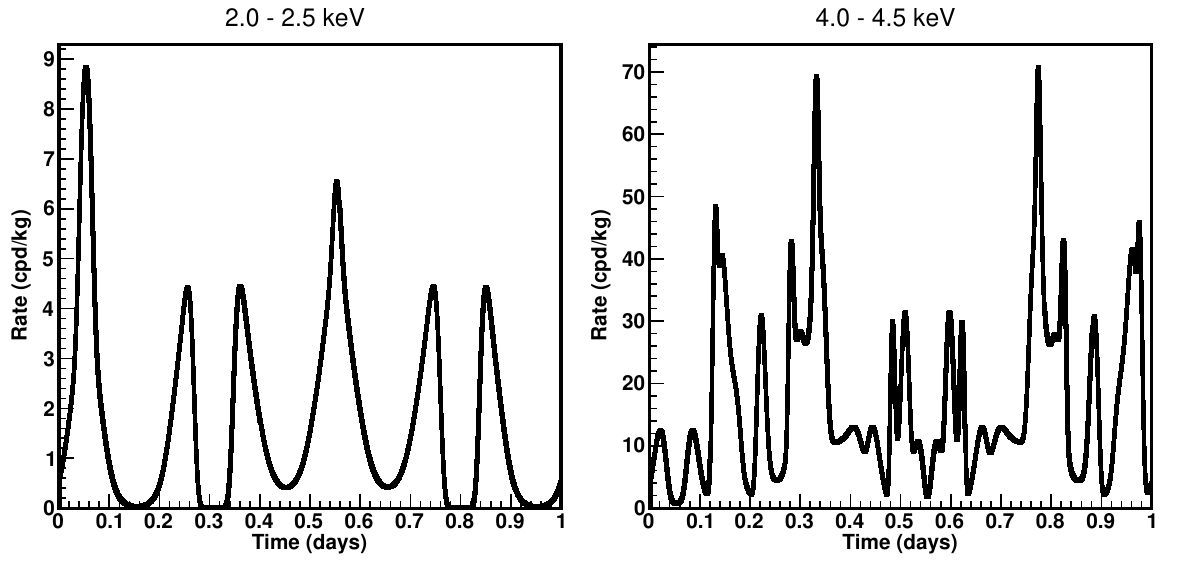}
\caption{
(a) Theoretical prediction of the count rate of photons converted from axions incident at a Bragg angle for a detector located at CJPL, China ($28.2^\circ$ N latitude, $101.7^\circ$ E longitude, and an altitude of 1580 m above sea level), assuming $g_{A\gamma}=10^{-8}$ GeV$^{-1}$ and energy resolution FWHM=0.195 keV at 10.37 keV. 
(b) The typical axion-photon conversion rates $R(E,t)$ for 2.0$-$2.5 keV and 4.0$-$4.5 keV energy bands . The time scale is from 0.0 to 1.0 days for both bands.
}
\label{fig:expect_count_rate}
\end{figure}

Figure~\ref{fig:expect_count_rate} depicts both time and energy variation of the theoretical prediction rate of photons converted from axions through Bragg diffraction, assuming the CDEX-1B detector to be located at CJPL, China ($28.2^\circ$ N latitude, $101.7^\circ$ E longitude, and an altitude of 1580 m above sea level with respect to the mean Earth ellipsoid) on April 1st, 2015, and the [001] axis of the crystal normally aligned with the vertical axis of the cryostat. 
The trajectory of the Sun on an arbitrarily chosen day is obtained from the Solar Position Algorithm (SPA) developed by the National Renewable Energy Laboratory (NREL)~\cite{REDA2004577_SPA}. The axion probability density function (PDF) is evaluated over the three-year period with five-minute precision.
The pronounced variation of the rate with daytime provides powerful background discrimination. 

\begin{figure}[!htbp]
\includegraphics[width=\linewidth]{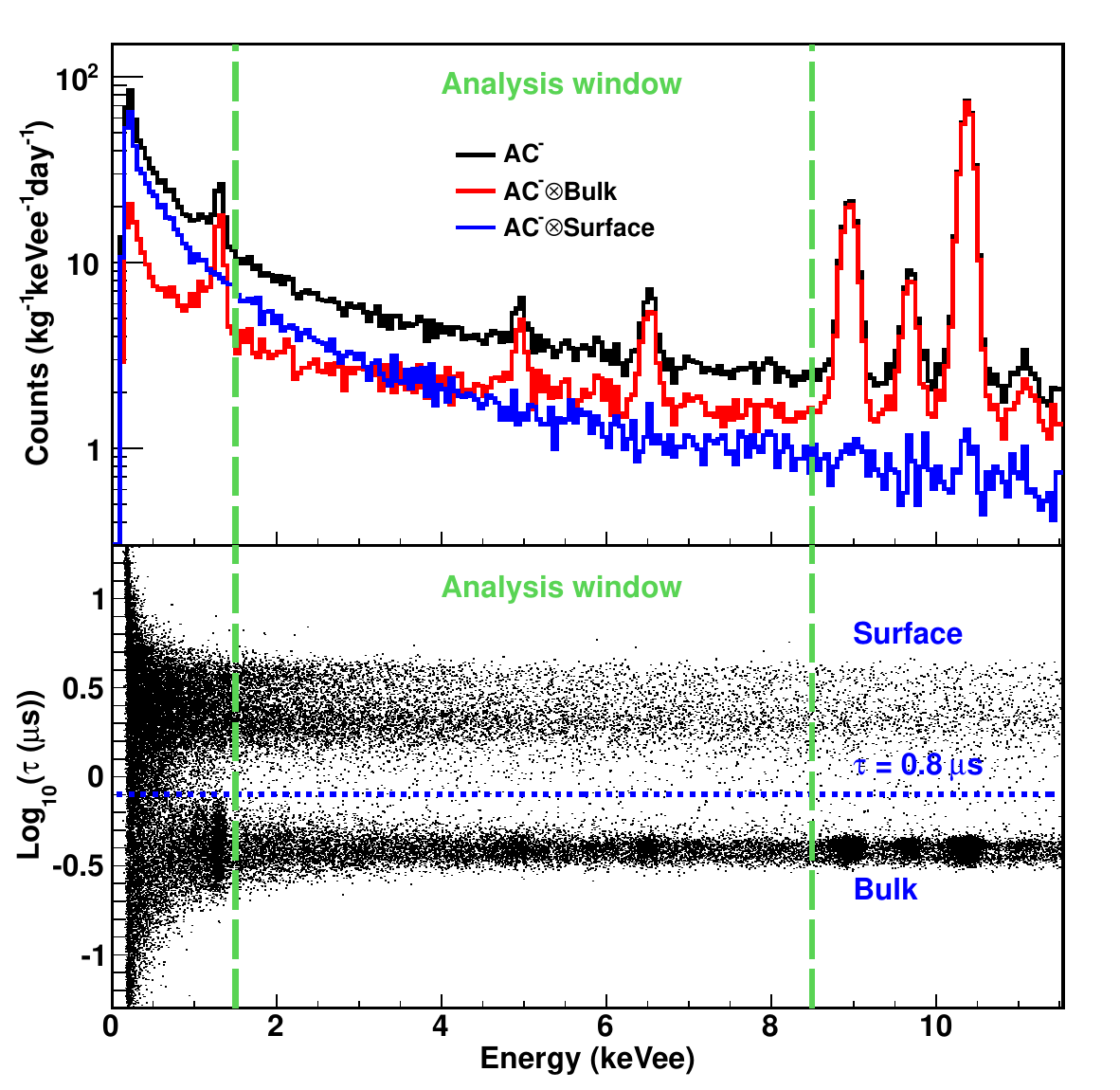}
\caption{
The analysis window, defined from 2.0$-$8.5 keV, is determined by the expected axion flux, background rate and detection efficiency.
}
\label{fig:bsplot}
\end{figure}

\section{III. DATA ANALYSIS}
\subsection{A. Candidate Event Selection}

The background spectrum is derived by the following procedures described in our earlier work~\cite{CDEX_1kg_2014,CDEX_1kg_2016,Yang:2018b}. 

(i) Energy calibration: The optimal area from $S_{p6}$ is selected to define the energy for its excellent energy linearity at the low energy region. Energy calibration was made with the internal cosmogenic x-ray peaks:  $^{68}$Ge (1.30 and 10.37 keV), $^{68}$Ga (9.66 keV), $^{65}$Zn (8.98 keV), and the zero energy defined by the RT events. 

(ii) Stability check, which discards the time periods of calibration or other testing experiments.

(iii) Anti-Compton veto, which removes the events in coincidence with the anti-Compton detector.

(iv) Physics vs. electronic noise, which discriminates physical events from electronic noise and spurious signals. 

(v) Bulk and surface event selection (BS): the events depositing energy in the surface layer with their characteristic slower rise-time will be rejected, due to the partial charge collection. The pulse shape analysis method, called ``ratio method'' based on the rise-time distribution PDFs (probability density functions), is applied to discard the surface events and derive the signal-retaining and background-leakage efficiencies~\cite{Yang:2018a}.

Figure~\ref{fig:bsplot}(a) shows the trigger efficiency and the variations of combined efficiencies with energy, including those from the trigger,  electronic noise events  and anti-Compton vetos efficiencies. The former two efficiencies are determined  by the survival of anti-Compton tag events (AC$^+$) from source samples and \emph{in situ} background, and the latter one is derived by the survival of RT events. 
As shown in Fig.~\ref{fig:bsplot}(b), the analysis window of this work  is chosen to be 2.0$-$8.5 keV, which contains most of the expected solar axion events. The combined efficiency is $17 \%$, and bulk/surface events shown in Fig.~\ref{fig:bsplot}(b) are well separated among that energy range. The summed background rate after correcting for detection efficiency is $\sim2.0$ cpkkd (counts per keV per kg per day).

\subsection{B. Analysis method}
The analysis method, as described in Ref.~\cite{CRESWICK:PLB1998,SOLAX:PRL1998}, is used to get the constraints of the axion-photon coupling $g_{A\gamma}$. The time correlation function is defined as follows:
\begin{equation}
\label{eq:chi}
\begin{aligned}
\chi = \int_0^T[R(t, E)-\bar{R}(E)]n(t)dt,
\end{aligned}
\end{equation}
where $R(t,E)$ defined in Eq.~\ref{eq:chi} is the theoretical axion count rate in a given energy $E$ and instantaneous time $t$. $T$ is the total period of data taking. $\bar{R}(E)$ is the average of $R(t,E)$ over total time period and $n(t)$ is the event count number at time $t$ in a short time interval. Note that if no solar axion events are observed, i.e., $n(t)$ is uncorrelated with the position of the Sun, then the average of $\chi$ converges to zero. Otherwise, it will increase proportionally to $T$.

In our analysis procedures, we divide the acquisition time $T$ and energy window 2.0$-$8.5 keV into $n$ pieces and $k$ pieces, respectively. The energy interval $\Delta E$ is set to 0.5 keVee, determined by the energy resolution. Then the observed $\chi_k^{\text{obs}}$ at energy interval ($E_k, E_k+\Delta E$) is re-written as below:
\begin{equation}
\begin{aligned}
\chi_k^ {\text{obs}}= \epsilon_k \sum\limits_{i} \left[ \overline{R_k}(t_i) - \left \langle \overline{R_k} \right \rangle\right] \cdot n_{ik}^{\text{obs}}\equiv \epsilon_k \sum\limits_{i}^{n} {W_{ik}\cdot n_{ik}^{\text{obs}}},
\end{aligned}
\end{equation}
where $\epsilon_k$ is the average detector efficiency at the energy interval $k$, $\overline{R_k}(t_i)$ represents the average expected rate at the time interval $(t_i, t_i+\Delta t)$ and $\left \langle \overline{R_k} \right \rangle$ indicates the average over the total time $T$. 
$n_{ik}^{\text{obs}}$ is the observed event number at time interval $t_i$ and energy interval $E_k$, whose average theoretical expectation $n_{ik}$ is to be composed of axion events and background events.

\begin{equation}
\begin{aligned}
n_{ik}= \epsilon_k \left[ \lambda\overline{R_k}(t_i)+b_k \right] \Delta t \Delta E,
\end{aligned}
\end{equation}
where $b_k$ is the background component that is constant in time and $\lambda\equiv(g_{A\gamma}\times10^{8}$ GeV)$^4$ is a dimensionless coupling (i.e., $\lambda=1$ equivalent to $g_{A\gamma}=10^{-8}$ GeV$^{-1}$). Therefore, the expected average value of $\chi_k$ is then
\begin{equation}
\begin{aligned}
\chi_k &= \epsilon_k \sum\limits_{i} \left[ \overline{R_k}(t_i) - \left \langle \overline{R_k} \right \rangle\right] \cdot n_{ik}\\
&\equiv \epsilon_k^2 \sum\limits_{i}^{n} {W_{ik} \cdot  \left[ \lambda\overline{R_k}(t_i)+b_k \right] \Delta t \Delta E}.
\end{aligned}
\end{equation}

Assuming $n_{ik}$ is dominated by the background component $b_k$, the expecited average and variance of $\chi_k$ will be written~\cite{Armengaud:JCAP2013}.
\begin{equation}
\begin{aligned}
& \left \langle \chi_k \right \rangle = \lambda \cdot \epsilon_k^2 \sum\limits_i{W_{ik}}^2 \Delta t \Delta{E} \equiv \lambda\cdot A_k,\\
& \sigma^2(\chi_k) \approx \epsilon_k b_k / A_k.
\end{aligned}
\end{equation}

In this analysis, the likelihood function is constructed as follows:
\begin{equation}
\begin{aligned}
 \mathcal{L}(\lambda) = \prod_k \exp \left[ \frac{-(\chi_k^{\text{obs}}-\chi_k(\lambda))^2}{2\sigma^2(\chi_k)} \right].
\end{aligned}
\end{equation}

Maximizing $\mathcal{L}$ in relation to $\lambda$ yields the maximum-likelihood estimator $\hat \lambda$,
\begin{equation}
\begin{aligned}
\hat \lambda = \frac{\sum_{k}{\chi_k^{\text{obs}}/(\epsilon_k b_k)}}{\sum_{k}{A_k/(\epsilon_k b_k)}}.
\end{aligned}
\label{eq:lambda}
\end{equation}

The estimator of the variance of $\lambda$ is also given
\begin{equation}
\begin{aligned}
\hat \sigma(\lambda) = \left(\sum_{k}\frac{A_k}{\epsilon_kb_k}\right)^{-1/2}.
\end{aligned}
\label{eq:err_lambda}
\end{equation}

\section{IV. RESULTS}
\begin{figure}[!htbp] 
\includegraphics[width=\linewidth]{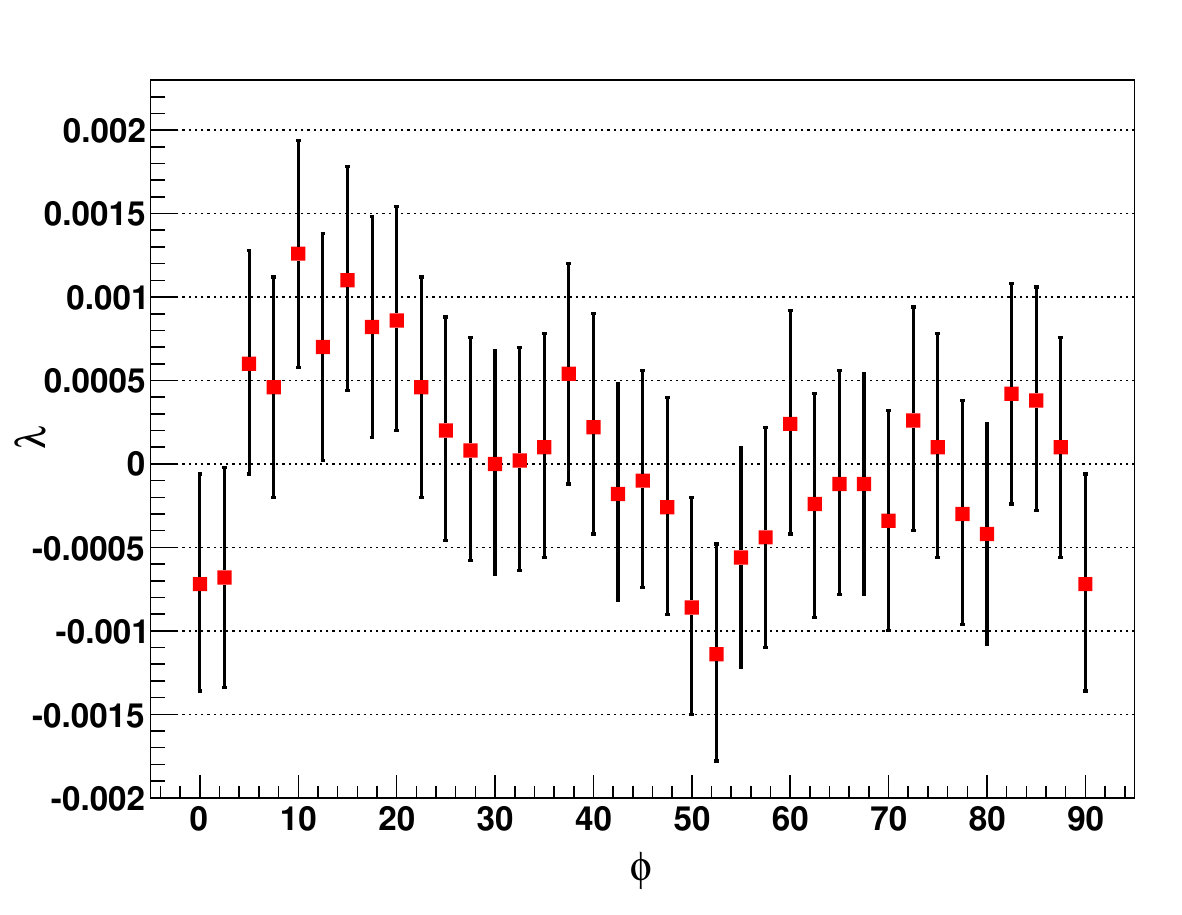}
\caption{
Values of $\lambda$ calculated from the CDEX-1B dataset as a function of the assumed azimuthal angle $\phi$ around the fixed vertical axis. The error bars are 1$\sigma$. The maximum of $\lambda$ is reached for $\phi=10.0$ degrees. The results are compatible with zero signal.
}
\label{fig:residualspec}
\end{figure}

Since the crystal was grown along the [100] axis, the orientation of the [100] axis has been determined. However, the orientations of the [010] and [001] axes are currently unknown, and the azimuthal orientation $\phi$ of the crystal has not been measured. Therefore, scan the different azimuthal angles $\phi$ and select the weakest bound to derive the upper limits of $\lambda$ or equivalently $g_{A\gamma}$.
Figure~\ref{fig:residualspec} shows the results for 1107.5 kg-days of data in the energy range of 2.0$-$8.5 keV in 0.5 keV intervals at various azimuthal angle $\phi$. Each data point of $\lambda$ in Fig.~\ref{fig:residualspec} and its corresponding error are defined by Eq.~\ref{eq:lambda} and Eq.~\ref{eq:err_lambda}.
All of the maximizations $\lambda$ are compatible with zero, and there is no evidence of solar axion conversion to photons.

The conservative upper limit with a 95\% C.L. for the axion-photon coupling among at all the azimuthal angles is:
\begin{equation}
\begin{aligned}
g_{A\gamma}<2.08\times10^{-9}~{\rm{GeV}}^{-1}.
\end{aligned}
\end{equation}

\begin{figure}[!htbp] 
\includegraphics[width=\linewidth]{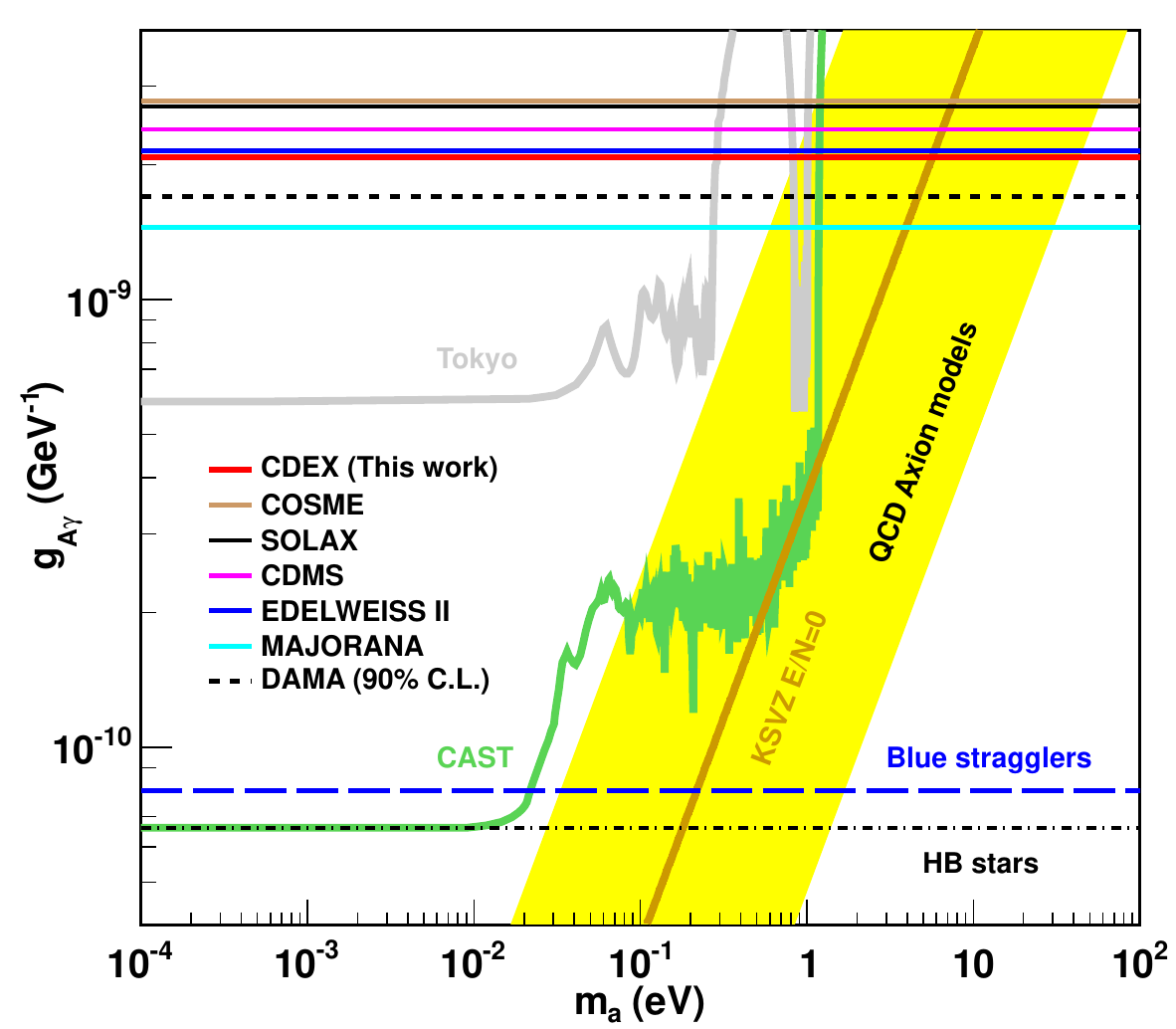}
\caption{
Exclusion plots on the $g_{A\gamma}$ vs axion-mass, as well as the results from other experiments~\cite{SOLAX:PRL1998,MORALES:AstroPhys2002,CDMS_PRL2009,Armengaud:JCAP2013,DAMA:PLB2001,Majorana_PRL2022}, CAST limits~\cite{Arik-CAST:PRL2014,CAST:Nature2017}, Tokyo limits~\cite{Tokyo:MORIYAMA1998,Tokyo:INOUE2002,Tokyo:INOUE2008} and astrophysical bounds~\cite{Raffelt2008,BlueStragglers_PRL2013,HDM_Hannestad_2010,HBconstrains2014,KSVZ_Luzio2017}. All limits are for 95\% CL, except for the 90\% DAMA limit. The KSVZ axion phase space is shown with the realistic range of E/N found in Ref.~\cite{KSVZ_Luzio2017}. 
}
\label{fig:exclusionplot}
\end{figure}

Figure~\ref{fig:exclusionplot} shows our exclusion curve with a 95\% C.L., along with the results from other experiments ~\cite{SOLAX:PRL1998,MORALES:AstroPhys2002,CDMS_PRL2009,Armengaud:JCAP2013,DAMA:PLB2001,Arik-CAST:PRL2014,CAST:Nature2017}, astrophysical bounds~\cite{Raffelt2008,BlueStragglers_PRL2013,HDM_Hannestad_2010,HBconstrains2014,KSVZ_Luzio2017}, and the benchmark DFSZ and KSVZ models~\cite{KSVZ_Luzio2017}. Our result surpasses over the limit from EDELWEISS-II experiment~\cite{Armengaud:JCAP2013}.

\section{V. Summary AND Prospects}

Based on the 1107.5 kg-days data from the CDEX-1B experiment, an analysis was performed to search the axion-photon conversion via Primakoff and the Bragg diffraction effect in germanium crystal, the rate of which is found to be consistent with zero and a new limit on the axion-photon coupling is obtained.
A competitive result among the germanium-based experiment has been achieved. It also constrains axion models in the mass range of~1$-$100 eV/$c^2$ for hadronic axions, complementing helioscopes experiment such as CAST~\cite{Arik-CAST:PRL2014,CAST:Nature2017} in the high mass region.

CDEX-50, the next generation of the CDEX experiment, is currently in preparation. CDEX-50 will use an array of 50 1-kg HPGe detectors with optimized electronics and will be operated in a superior radioactive environment~\cite{CDEX_DMe,CDEX50}. Improvements are being made to the accurate measurement of all crystal orientations, the reduction of the radiation background, and the homemade germanium detectors with ultralow-background electronics. Both enhancements to the detector exposure and the background level (0.01 cpkkd) will promote the next order of sensitivity.

This work was supported by the National Key Research and Development Program of China (Grants No. 2023YFA1607100 and No. 2022YFA1605000), the National Natural Science Foundation of China (Grants No. 12322511, No. 12175112, No. 12005111, and No. 11725522), and the Sichuan Provincial Natural Science Foundation (No. 2022NSFSC1825). We acknowledge the Center of High-performance Computing, Tsinghua University, for providing the facility support. We would like to thank CJPL and its staff for hosting and supporting the CDEX project. CJPL is jointly operated by Tsinghua University and Yalong River Hydropower Development Company.

\bibliography{CDEX1B_Axion_Paper.bib}

\end{document}